\documentclass[10pt]{article}
\usepackage{amsmath,verbatim,algorithmic,algorithm,graphicx}
\begin{document}

\begin{title}
{SLIQ: Simple Linear Inequalities for Efficient Contig Scaffolding}
\end{title}
\date{}
\author{
Rajat S. Roy$^{ \ast}$,
Kevin C. Chen,
Anirvan M. Sengupta,
Alexander Schliep}
\maketitle

\begin{abstract}
Scaffolding is an important subproblem in {\it de novo} genome assembly 
in which mate pair data are used to construct a linear sequence of 
contigs separated by gaps. Here we present SLIQ, a set of simple 
linear inequalities derived from the geometry of contigs on the line that can
be used to predict the relative positions and orientations of contigs from individual
mate pair reads and thus produce a contig digraph. 
The SLIQ inequalities can also filter out 
unreliable mate pairs and can be used as a preprocessing step for any
scaffolding algorithm. We tested the SLIQ inequalities on five real data 
sets ranging in complexity from simple bacterial genomes to complex mammalian 
genomes and compared the results to the majority voting procedure used by many
other scaffolding algorithms. SLIQ predicted the relative positions and 
orientations of the contigs with high accuracy in all cases and
gave more accurate position predictions than majority voting for
complex genomes, in particular the human genome. 
Finally, we present a simple scaffolding 
algorithm that produces linear scaffolds given a contig digraph. We show 
that our algorithm is very efficient compared to other scaffolding algorithms 
while maintaining high accuracy in predicting both contig positions and 
orientations for real data sets.

\end{abstract}
\section{Introduction}
{\it De novo} genome assembly is a classical problem in bioinformatics in 
which short DNA sequence reads are assembled into longer blocks of contiguous 
sequence (contigs) which are then arranged into linear chains of contigs 
separated by gaps (scaffolds). Modern genome sequencing projects typically 
include mate pair reads in which the approximate distance between a pair of 
reads plus the two read lengths (the insert length) is fixed during the experimental construction of 
the sequencing library. Some genome projects also include mate pair libraries
with several different insert lengths. Although there are experimental 
differences between mate pairs and paired-end reads, we will refer to
them interchangeably as mate pairs in this paper since we can treat them
identically from an algorithmic point of view. 

Computational genome assembly is typically performed in at least two stages --- 
the contig 
building stage and the scaffolding stage. In this paper we do not address
the contig building problem but rather assume that we have access to a set of
contigs produced by an independent algorithm. However we discuss the 
relationship of the contig building and scaffolding stages later in the 
Discussion.
For the scaffolding problem, the most popular strategy is to construct the 
contig graph in which nodes represent contigs and edges represent
sets of mate pairs connecting two contigs (i.e. the two reads of the mate 
pair fall in the two different contigs). The edges are given weights 
equal to the number of mate pairs connecting the two contigs.
 
A common procedure is to filter out unreliable edges by picking a small 
threshold (commonly 2-5) and removing all edges with weight less than that 
threshold. For the remaining edges, a majority vote is used to decide on the 
relative orientation and position of the contigs. This simple majority 
voting strategy is implemented in a number of commonly-used assemblers and 
stand-alone scaffolders including ARACHNE ~\cite{arachne}, BAMBUS~\cite{bambus}, 
SOPRA~\cite{sopra} and SOAPdenovo~\cite{soap_dn} with various choices of
threshold. Opera~\cite{opera} and the Greedy Path-Merging 
algorithm~\cite{path_merge} use a different strategy to bundle edges. Given 
a set of mate pairs connecting two contigs, these algorithms calculate the 
median and standard deviation of the insert lengths of the set of mate pairs 
and create a bundle using only 
mate pairs with insert length that are close to the median. 
ALLPATHS~\cite{allpaths1} and VELVET~\cite{velvet} do not build the
contig graph and thus do not have a read filtering step similar to the 
other assemblers mentioned. 
The majority voting procedure implicitly assumes that 
misleading mate pairs are random and independently generated and that 
majority voting should eliminate the problematic mate pairs. However, this 
assumption is often not true because of the complex repeat structure of 
large genomes, such as human. 

In this paper, we show that unreliable mate pairs can be reliably 
filtered using SLIQ, a set of simple linear inequalities derived from the
 geometry of contigs on the line.
Thus SLIQ produces a reduced subset of reliable mate pairs 
and thus a sparser graph which results in a simpler optimization problem for
the scaffolding algorithm. More importantly, SLIQ can be used to predict the 
relative positions and orientations of the contigs, yielding a {\em directed} contig graph. 
Our experiments show that both SLIQ and majority voting are very accurate at 
predicting relative orientations but SLIQ is clearly more accurate at 
predicting relative positions for complex genomes.

The simplicity of SLIQ makes it very easy to integrate as a preprocessing step
to any existing scaffolders, including recent scaffolders such as MIP scaffolder~\cite{mips},
Bambus 2~\cite{bambus2} and SSPACE~\cite{sspace}.
 To illustrate the effectiveness of SLIQ, we implemented a naive 
scaffolding algorithm that produces linear scaffolds from the contig digraph. 
We show that despite its simplicity, our naive scaffolder provides 
very accurate draft scaffolds, comparable to or improving upon the more complicated
sate of the art, very quickly. 
These scaffolds can either be output directly or
used as reasonable starting points for further refinement with more complex scaffolding algorithms.

\section{Algorithms}
\begin{figure}
\setlength{\unitlength}{0.04cm}
\begin{center}
\caption{The geometry of two contigs, $C_i$ and $C_j$, arranged on a line with
relevant quantities indicated.}
\label{gaps_fig}
\setlength{\unitlength}{0.04cm}
\begin{picture}(500,60)
\linethickness{0.3mm}
\put(20,30){\vector(1,0){150}}
\put(200,30){\vector(1,0){200}}
\put(75,35){$C_i$}
\put(300,35){$C_j$}

\put(10,25){$P_i$}
\put(188,25){$P_j$}
\linethickness{0.1mm}
\put(140,40){\line(1,0){85}}
\put(170,42){$L$}
\put(140,30){\line(0,1){10}}
\put(225,30){\line(0,1){10}}
\linethickness{0.2mm}
\put(140,32){\vector(1,0){20}}
\put(160,30){\line(0,1){4}}
\put(225,32){\vector(-1,0){20}}
\put(205,30){\line(0,1){4}}
\put(137,23){$o_i$}
\put(202,23){$o_j$}
\put(222,22){$o_j+R$}
\linethickness{0.1mm}
\put(170,5){\line(0,1){30}}
\put(200,5){\line(0,1){30}}
\put(192,15){\line(1,0){8}}
\put(178,15){\line(-1,0){8}}
\put (180,12){$g_{ij}$}
\put(20,30){\line(0,1){30}}
\put(400,30){\line(0,1){30}}
\put(20,55){\line(1,0){165}}
\put(210,55){\line(1,0){190}}
\put(188,55){$-g_{ji}$}

\put(20,30){\line(0,-1){30}}
\put(20,15){\line(1,0){70}}
\put(178,15){\line(-1,0){70}}
\put(100, 13){$l_i$}

\put(400,30){\line(0,-1){30}}
\put(200,15){\line(1,0){90}}
\put(400,15){\line(-1,0){90}}
\put(300, 13){$l_j$}

\end{picture}
\end{center}
\end {figure}
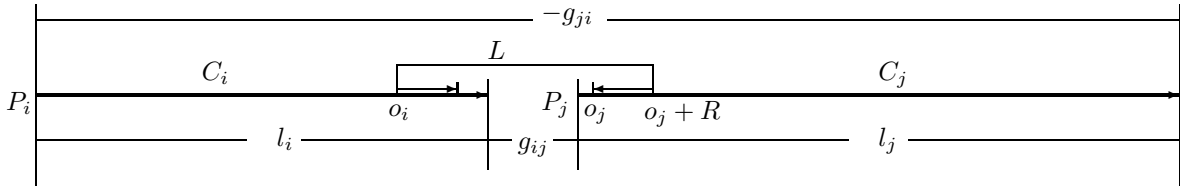

We begin with a high level outline of our algorithm for constructing a 
directed 
contig graph (Algorithm \ref{outline_alg}). The crux of the algorithm is 
SLIQ, a set of simple linear inequalities that are used to filter mate pairs
and predict the relative position and orientation of contigs. 
In subsequent sections, we will present proofs for the SLIQ 
inequalities and a detailed version of the digraph construction algorithm 
(Algorithm \ref{alg1}). Finally, we will present a simple scaffolding 
algorithm (Algorithm \ref{naive_alg}) that uses the contig digraph to 
construct draft scaffolds. Throughout the paper 
we will abbreviate mate pair reads as {\it MPR}.
\begin{algorithm}
\caption{Construct Contig Digraph (Outline)}
\label{outline_alg}
\begin{algorithmic}[1]
\REQUIRE {\it input:} $P=$ a set of MPRs that connect two contigs, $C=$ a set of contigs
\STATE Construct the contig graph $G$ with vertex set $C$ and edges 
representing MPRs from $P$ that pass a certain majority cutoff.
\STATE Find a good orientation assignment for the contigs ($\Theta=\{\Theta_1, \Theta_2,\ldots\}$) where $\Theta_i$ is the orientation of the $i$th contig,
for example by finding a spanning tree of $G$. 
\STATE Define $M_p$ to be the set of MPRs that satisfy the SLIQ inequalities
\STATE Construct a directed contig graph $G_d$ with vertex set $C$ and edges
representing MPRs from $M_p$ that pass certain criteria.
\end{algorithmic}
\end{algorithm}

\subsection{Definitions and Assumptions}
For the sake of deriving the SLIQ inequalities, we assume that  
we know the position of the contigs on the reference genome. However, this 
information cancels out later on which allows us to analyze the MPRs 
without access to prior contig position information. For the derivation we 
also assume that all the contigs have the same orientation. Later we will not 
need this information. 

Let $P_i$ be the position of contig $C_i$ 
in the genome and $l_i$ be the length of the contig (Fig. \ref{gaps_fig}). 
We define gap $g_{ij}$ to be the difference between the start 
position of contig $C_j$ and the end position of contig $C_i$, and similarly
for $g_{ji}$: 

\begin{align}\label{gap_def}
g_{ij}=P_j- P_i-l_i, \\
g_{ji}=P_i- P_j-l_j.\nonumber
\end{align}
We assume that the maximum overlap of two contigs 
is one read length, $R$. In practical contig building
software based on De Bruijn graphs, the maximum overlap is usually one 
$k$-mer where $R > k$ so our assumption is valid.

\subsection{Derivation of Two Gap Equations}

If we assume that $P_i < P_j$ as in Fig. \ref{gaps_fig} and that the maximum 
overlap between two contigs is $R$  
(i.e. the minimum gap $g_{ij}$ is $-R$), then\\
\begin{align}
P_j- P_i-l_i &\geq -R,\nonumber \\
P_j- P_i &\geq l_i-R.
\end{align}
Now consider the quantity $g_{ij} - g_{ji}$. 
Using \eqref{gap_def}, we can derive the following inequality which 
we call Gap Equation 1\\
\begin{align}\label{gap_dif}
g_{ij}-g_{ji}&=2\,(P_j- P_i)+(l_j-l_i)\nonumber \\
&\geq 2l_i-2R+l_j-l_i\nonumber \\
&\geq l_i+l_j -2R.
\end{align}
Therefore, we have shown that 
$(P_i < P_j )\Rightarrow (g_{ij}-g_{ji}\geq l_i+l_j -2R)$.
Next consider the quantity $g_{ij} + g_{ji}$. We can easily derive Gap
Equation 2:
\begin{align}\label{gap_sum}
g_{ij}+g_{ji}&=-(l_j+l_i).
\end{align}

Now we will prove the other direction of the implication in Gap Equation 1 
and show that $(g_{ij}-g_{ji}\geq l_i+l_j -2R)\Rightarrow (P_i < P_j )$.
Using Gap Equation 1 and Equation \eqref{gap_def}, we get
\begin{align}\label{gap_ineq}
g_{ij}-g_{ji}&\geq l_i+l_j -2R,\nonumber \\
2(P_j-P_i)+(l_j-l_i)&\geq l_i+l_j -2R,\nonumber \\
2(P_j-P_i)&\geq 2l_i -2R,\nonumber \\
P_j-P_i&\geq l_i -R.\nonumber \\
\end{align}
Since no contig length can be less than $R$, the length of a read,  
$l_i-R > 0$ and hence, $P_j-P_i>0$ or $P_i<P_j$.
Therefore, $(g_{ij}-g_{ji}\geq l_i+l_j -2R)\Rightarrow (P_i < P_j )$
and together we have proven,
\begin{equation}\label{gap_equiv}
(g_{ij}-g_{ji}\geq l_i+l_j -2R)\iff (P_i < P_j ).
\end{equation}

\subsection{Using the Gap Equations to Predict Relative Positions}
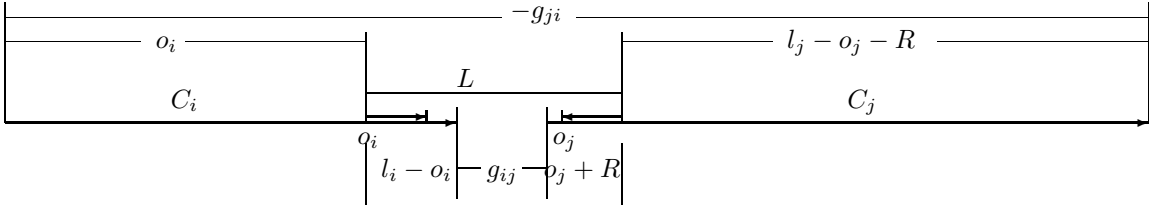
\begin{figure}
\setlength{\unitlength}{0.04cm}
\begin{center}
\caption{The geometry of two contigs arranged on a line in terms of quantities
 known in {\it de novo} assembly.}
\label{gaps_fig_2}
\begin{picture}(520,100)
\linethickness{0.3mm}
\put(20,30){\vector(1,0){150}}
\put(200,30){\vector(1,0){200}}
\put(75,35){$C_i$}
\put(300,35){$C_j$}
\linethickness{0.1mm}
\put(140,40){\line(1,0){85}}
\put(170,42){$L$}
\put(140,30){\line(0,1){10}}
\put(225,30){\line(0,1){10}}
\linethickness{0.2mm}
\put(140,32){\vector(1,0){20}}
\put(160,30){\line(0,1){4}}
\put(225,32){\vector(-1,0){20}}
\put(205,30){\line(0,1){4}}
\put(137,23){$o_i$}
\put(202,23){$o_j$}
\linethickness{0.1mm}
\put(170,5){\line(0,1){30}}
\put(200,5){\line(0,1){30}}
\put(192,15){\line(1,0){8}}
\put(178,15){\line(-1,0){8}}
\put (180,12){$g_{ij}$}
\put(140,23){\line(0,-1){20}}
\put(225,23){\line(0,-1){20}}
\put(145,12){$l_i-o_i$}
\put(199,12){$o_j+R$}
\put(20,30){\line(0,1){40}}
\put(400,30){\line(0,1){40}}
\put(140,30){\line(0,1){30}}
\put(225,30){\line(0,1){30}}
\put(70,55){$o_i$}
\put(280,55){$l_j-o_j-R$}
\linethickness{0.05mm}
\put(20,57){\line(1,0){45}}
\put(80,57){\line(1,0){60}}
\put(225,57){\line(1,0){50}}
\put(400,57){\line(-1,0){70}}
\put(20,65){\line(1,0){165}}
\put(210,65){\line(1,0){190}}
\put(188,65){$-g_{ji}$}
\end{picture}
\end{center}
\end{figure}

Our definitions in Equation \eqref{gap_def} used the quantities
$P_i$ and $ P_j$ which are 
not available in practice in {\it de novo} assembly. Thus we need to define 
the gaps $g_{ij}$ and $g_{ji}$ in terms of quantities we know such as the 
insert length $L$ and the read offsets relative to the contigs $o_i$ and 
$o_j$. Note that the insert length for each MPR is an unknown constant so 
treating it as a constant in the proof is justified. In practice, we use 
$L=\bar{L}+2\sigma$, where $\bar{L}$ is the reported or computed mean 
and $\sigma$ is the standard deviation of the insert length distribution.

Let $L$ be the insert length, $o_i$ and $o_j$ be the offsets
of the start positions of the paired reads in $C_i$ and $C_j$ respectively and 
$\Theta_i$ and $\Theta_j$ be the
orientations of $C_i$ and $C_j$ respectively. To simplify the notation we 
abbreviate $\Theta_i= \Theta_j$ as $\Theta_{i=j} $
and $\Theta_i\neq \Theta_j$ as $\Theta_{i\neq j}$. Then,  if $P_i < P_j$ and $\Theta_{i=j}$ (see Fig. \ref{gaps_fig_2}), 
we can redefine the gaps $g_{ij}$ and $g_{ji}$ without using the contig start positions $P_i$ and $P_j$:
\begin{align}
g_{ij}&=L-l_i+o_i-o_j-R \label{gap_off_1},\\
g_{ji}&=-L -l_j+o_j+R-o_i\label{gap_off_2}.
\end{align}
Note that these definitions remain consistent with Gap Equation 2 
(Equation \eqref{gap_sum}). 
Taking the difference 
of Equations \eqref{gap_off_1} and \eqref{gap_off_2} we can similarly
remove $P_i$ and $P_j$ from Gap Equation 1:\\
\begin{equation}\label{off_dif}
g_{ij}-g_{ji}=2L-2R+2(o_i-o_j)+(l_j-l_i).
\end{equation}
Using Equations \eqref{off_dif} and \eqref{gap_ineq}, we derive the following
inequality:  \\
\begin{align}
2L-2R+2(o_i-o_j)+(l_j-l_i)&\geq l_i+l_j -2R,\nonumber \\
2L+2(o_i-o_j)+(l_j-l_i)&\geq l_i+l_j, \nonumber \\
L+(o_i-o_j)&\geq l_i.\nonumber
\end{align}
Consequently we obtain that
 $(P_i < P_j )\wedge\Theta_{i=j}\Rightarrow L+(o_i-o_j)\geq l_i$. 
Negating the implication gives \\
\begin{align*}
\neg ( L+(o_i-o_j)\geq l_i) &\Rightarrow \neg ((P_i < P_j )\wedge\Theta_{i=j}),\\
L+(o_i-o_j)< l_i&\Rightarrow (P_i> P_j )\vee \Theta_{i\neq j}.
\end{align*}
Now without loss of generality we can assume that $\Theta_{i\neq j}$ is false.
This is possible because our experiments later show that the SLIQ or majority 
voting procedures are both very accurate at predicting relative orientation 
(Table \ref{edge_results}) so we can first determine the relative orientations 
of the contigs and flip the orientation of one contig if required. 
Thus we have \\
\begin{equation}\label{off_imp}
L+(o_i-o_j)< l_i\Rightarrow (P_i> P_j ).
\end{equation}
In addition, we introduce two filters that are very useful in practice for
removing unreliable MPRs. To derive the first filter, if $P_j<P_i$, 
\begin{align}\label{off_l_bound}
L&=l_j-o_j+g_{ji}+o_i+R, \nonumber \\
&\geq l_j-o_j-R+o_i+R,\nonumber \\
o_j-o_i&\geq l_j -L,\nonumber \\
o_i-o_j&< - l_j +L.
\end{align}
The second filter is to discard an MPR  
if it passes the test for both $P_i<P_j$ and $P_j<P_i$.
\subsection{Using the Gap Equations to Predict Relative Orientations}
So far we have only predicted relative positions when $\Theta_{i=j}$.
Now we show that we can also use the gap equations to infer the relative
orientations of the contigs.
First, if $(P_i < P_j )$ and the minimum gap is $-R$ then we have
\begin {equation}\label{ori1}
g_{ij}=L-l_i+o_i-o_j-R\geq -R.
\end{equation}
Similarly, if $(P_j < P_i )$, then we define $\bar{g}_{ji}$ and write
\begin {equation}\label{ori2} 
\bar{g}_{ji}=L-l_j+o_j-o_i-R\geq -R.
\end{equation}
Note that $\bar{g}_{ji}$ is different than $g_{ji}$ which we defined under
the assumption $P_i<P_j$ in Equation \eqref{gap_off_2}.

Since $(P_i < P_j )$ and $(P_j < P_i )$ are mutually exclusive and exhaustive neglecting $P_i=P_j$, 
at least one of Equations \eqref{ori1} 
and \eqref{ori2} will be true. Note that possibly also both could be true. For example,
if $P_i < P_j$ then $g_{ij}\geq -R$. Now $(P_j < P_i )$ must be false, but 
that does not imply that $\bar{g}_{ji}\geq -R$ is false. 
If both Equations \eqref{ori1} and \eqref{ori2} are true, then we can add them
to get $2L\geq l_i+l_j$. To summarize,
\begin{align*}
\big((g_{ij}\geq -R)  \wedge (\bar{g}_{ji}\geq -R)\big) &\Rightarrow 2L\geq l_i+l_j,\\
2L< l_i+l_j &\Rightarrow \big(\neg(g_{ij}\geq -R)  \vee \neg(\bar{g}_{ji}\geq -R)\big) 
\end {align*}
Recalling again that at least one of Equations \eqref{ori1} and \eqref{ori2} 
are true, we see that $2L< l_i+l_j$ is a sufficient condition for mutual 
exclusion (the XOR relation is denoted by $\oplus$):\\
\begin {align}
\Theta_{i=j} \wedge (2L< l_i+l_j)&\Rightarrow (g_{ij}\geq -R ) \oplus (\bar{g}_{ji}\geq -R),\nonumber \\
\neg\big((g_{ij}\geq -R)  \oplus (\bar{g}_{ji}\geq -R)\big) &\Rightarrow \neg\big(\Theta_{i=j} \wedge (2L< l_i+l_j)\big),\nonumber \\
\neg\big((g_{ij}\geq -R)  \oplus (\bar{g}_{ji}\geq -R)\big) &\Rightarrow \big(\Theta_{i\neq j} \vee (2L\geq l_i+l_j)\big).\nonumber \\
\end{align}
If we use this equation only when the MPR and contigs satisfy the
inequality $2L< l_i+l_j$, we can then make the relative orientation prediction 
\begin {equation}\label{ori_imp}
\neg\big((g_{ij}\geq -R ) \oplus (\bar{g}_{ji}\geq -R)\big) \Rightarrow \Theta_{i\neq j} .
\end{equation}
Intuitively, the condition $2L< l_i+l_j$ means that the contig lengths should
be large relative to the insert length in order for the SLIQ method to work.
To find contigs of the same orientation, we arbitrarily flip one contig and 
run the above tests again, only this time if Equation \eqref{ori_imp} holds, 
then we conclude that the contigs were actually of the same orientation.
Say we flip $C_i$. We call the new offset $o_{\widehat{i}}$. Then 
\begin {align*}
\neg\big((g_{{\widehat{i}}j}\geq -R )\oplus (\bar{g}_{j\widehat{i}}\geq -R)\big) \Rightarrow \Theta_{{\widehat{i}}\neq j} \Rightarrow \Theta_{i=j}.
\end{align*}

Again, we introduce two additional filters that are very useful in
practical applications. First, if we find an MPR that
predicts both $\Theta_{i\neq j}$  and $\Theta_{i= j} $ then we leave it out of
consideration. Second, if the SLIQ equations imply $\Theta_{i\neq j}$, then we
require that both the reads of the MPR have the same mapping directions on 
the contigs and similarly for $\Theta_{i=j}$. 

We summarize our results in the following lemmas and Algorithm \ref{alg1}.
\newtheorem{lemma}{Lemma}
\begin{lemma}\label{pred_ori}
If the maximum overlap between contigs is $R$ and $2L< l_i+l_j$, then
\begin{center}
$\neg\big((g_{ij}\geq -R ) \oplus (\bar{g}_{ji}\geq -R)\big) \Rightarrow \Theta_{i\neq j},$\\
$\neg\big((g_{{\widehat{i}}j}\geq -R )\oplus (\bar{g}_{j\widehat{i}}\geq -R)\big) \Rightarrow \Theta_{i= j}. $
\end{center}
\end{lemma}
\begin{lemma}\label{pred_pos}
If the maximum overlap between contigs is $R$, the contigs have the same
orientation, (i.e. $\Theta_{i=j}$), then 
\begin{center}
$\big(L+(o_i-o_j)<l_i\big)\Rightarrow (P_i> P_j ).$
\end{center}
\end{lemma}
We also summarize the SLIQ inequalities,
\begin{align*}
g_{ij}-g_{ji}&\geq l_i-l_j-2R,\\
g_{ij}+g_{ji}&=-(l_j+l_i),\\
(g_{ij}-g_{ji}\geq l_i+l_j -2R)&\iff (P_i < P_j ),\\
g_{ij}-g_{ji}&=2L-2R+2(o_i-o_j)+(l_j-l_i).
\end{align*}
\begin{algorithm}
\caption{Construct Contig Digraph}
\label{alg1}
\begin{algorithmic}[1]
\REQUIRE {\it input:} $M=$ a set of MPRs connecting contigs, $C=$ a set of 
contigs, $w=$cutoff weight
\STATE Define $E'=\{(C_i,C_j):\text{an MPR connects } C_i$ and $C_j\}$
\STATE Let $wt(i,j)=$ (number of MPRs suggesting that $C_i$ and  $C_j$ have 
the same orientation) $-$ (number of MPRs suggesting that $C_i$ and $C_j$ have 
different orientations)
\STATE $E=\{(C_i,C_j):(i,j) \in E' \wedge wt(i,j)\geq w\}$
\STATE Construct a contig graph $G$ with vertex set $C$ and edge set $E$.
\STATE Find a good orientation assignment ($\Theta=\{\Theta_1, \Theta_2,\ldots\}$) for the contigs, for example, by finding a spanning tree of $G$. 
\STATE Set $M_p=\{\}$
\FORALL {$p: p\in M$}
\STATE Let $C_i$ and $C_j$ be the contigs connected by $p$.
\IF {$\Theta_{i=j}$}
	\IF {$\big(L+(o_i-o_j)<l_i\big)$ AND ($o_i-o_j < -l_i+L$)}
	\STATE predict $P_i> P_j$
	\STATE $M_p=M_p \cup \{p\}$
	\ENDIF
	 \IF {$\big(L+(o_j-o_i)<l_j\big)$ AND ($o_j-o_i < -l_j+L$)}
	\STATE predict $P_i< P_j$
	\STATE $M_p=M_p \cup \{p\}$
	\ENDIF
\ENDIF

\ENDFOR
\STATE Let $|E(i,j)|$ be the number of MPRs from $M_p$ that predict that 
$P_i<P_j$
\STATE Define $E_d=\{(C_i,C_j): |E(i,j)| > |E(j,i)|\}$
\STATE Output a contig digraph $G_d$ with vertex set $C$ and edge set $E_d$.
\end{algorithmic}
\end{algorithm}
\subsection{Illustrative Cases and Examples from Real Data}

\begin{figure}[ht]
\begin{center}
\caption{Illustrative cases in which both reads of the MPR fall in the center 
of the contigs (left) and the contigs have reversed positions (right).}
\label{mid_reads}
\begin{picture}(370,100)
\linethickness{0.3mm}
\put(20,30){\vector(1,0){50}}
\put(90,30){\vector(1,0){60}}

\linethickness{0.1mm}
\put(45,40){\line(1,0){75}}
\put(85,40){$L$}
\put(45,30){\line(0,1){10}}
\put(120,30){\line(0,1){10}}
\linethickness{0.2mm}
\put(45,32){\vector(1,0){20}}
\put(120,32){\vector(-1,0){20}}

\linethickness{0.3mm}
\put(190,30){\vector(1,0){100}}
\put(180,25){$P_i$}
\put(200,40){\vector(1,0){150}}
\put(190,40){$P_j$}
\linethickness{0.1mm}
\put(230,50){\line(1,0){75}}
\put(290,55){$L$}
\put(230,30){\line(0,1){20}}
\put(305,40){\line(0,1){10}}
\linethickness{0.2mm}
\put(250,32){\vector(-1,0){20}}
\put(285,42){\vector(1,0){20}}

\linethickness{0.1mm}
\put(200,40){\line(0,-1){30}}
\put(290,30){\line(0,-1){20}}
\put(200,15){\line(1,0){35}}
\put(290,15){\line(-1,0){35}}
\put(240,13){$g_{ij}$}

\put(190,30){\line(0,1){50}}
\put(350,40){\line(0,1){40}}
\put(190,75){\line(1,0){70}}
\put(350,75){\line(-1,0){70}}
\put(265,73){$g_{ji}$}
\end{picture}

\end{center}
\end {figure}
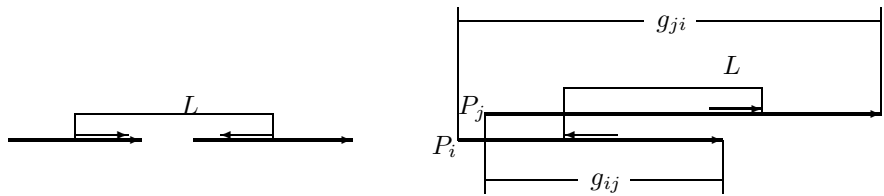

\label{two_cases}
In this section we present two illustrative cases that provide the 
intuition underlying the SLIQ equations. The ideal case for an MPR connecting 
two contigs is illustrated in Fig \ref{gaps_fig}. 
In that case the contigs are long compared to the insert length and the 
reads are mapped to the ends of the contigs. However, this situation does not 
always occur. Suppose the contigs are short such that the two reads of an MPR 
fall exactly in the center of the contigs. Then the right hand side of 
Equation \eqref{off_dif} reduces to $2L-2R$. So for both cases $P_i<P_j$ and 
$P_j<P_i$,
the right hand side of Equation \eqref{off_dif} has the same value, 
making it impossible to predict the relative positions of the two contigs. 
This situation is illustrated in Fig. \ref{mid_reads} on the left.
It is easy to see that prediction becomes easier as the contigs get longer
and the reads move away from the center of the contigs.

Now assume that the working assumption is $P_i<P_j$ but in reality, 
the reverse ($P_j<P_i$) is true. Then given that the contigs are long and 
reads map to the edges of the contigs, the insert length
$L$ would suggest the scenario depicted in Fig. \ref{mid_reads} (right side). 
This would make both $g_{ij}$ and $g_{ji}$ (as calculated
from Equations \eqref{gap_off_1} and \eqref{gap_off_2}) smaller than they
should be. In reality, the position of the contigs is similar to
that shown in Fig. \ref{gaps_fig} where we see that both $g_{ij}$ and $g_{ji}$
are larger than in Fig. \ref{mid_reads} (right side).
These wrong values would then be too small to satisfy 
the left hand side of Equation \eqref{gap_equiv} and this would demonstrate 
that the working assumption of $P_i<P_j$ is wrong.

\begin{figure}[ht]
\begin{center}
\caption{Three real examples of SLIQ predictions from the PSY dataset. 
For the correct prediction the equation $L+(o_i-o_j)< l_i$ evaluates to 
$3385< 5043$.
In the wrong prediction, it should have satisfied $L+(o_j-o_i)< l_j$  
but one of the contigs is smaller than the insert length
so it evaluates to $262<217$ (false). However 
$L+(o_i-o_j)< l_i$ evaluates to $498< 863$ so the 
wrong prediction is made. In the no prediction case, the condition $o_i-o_j< -l_j +L$ is violated. Even if that did not
fail, since one of the offsets falls almost in the center of a contig, both 
the conditions $L+(o_j-o_i)< l_j,  (299< 1384) $
and $L+(o_i-o_j)< l_i, (461< 506)$ are satisfied and we would not give a 
prediction for this MPR. To simplify the calculations we used $L=380$.}
\label{real_exm}
\begin{picture}(500,150)
\linethickness{0.3mm}
\put(30,90){\vector(1,0){64}}
\put(120,90){\vector(1,0){100}}
\put(50,80){$l_i=3149$}
\put(150,80){$l_j=5049$}
\put(75,110){Correct Prediction}
\linethickness{0.1mm}
\put(90,100){\line(1,0){35}}
\put(90,90){\line(0,1){10}}
\put(48,95){$o_i=3051$}
\put(125,90){\line(0,1){10}}
\put(130,95){$o_j=46$}

\linethickness{0.3mm}
\put(250,90){\vector(1,0){43}}
\put(300,90){\vector(1,0){150}}
\put(250,80){$l_j=217$}
\put(350,80){$l_i=863$}
\put(275,110){Wrong Prediction}
\linethickness{0.1mm}
\put(270,100){\line(1,0){73}}
\put(270,90){\line(0,1){10}}
\put(235,95){$o_j=99$}
\put(343,90){\line(0,1){10}}
\put(345,95){$o_i=217$}

\linethickness{0.3mm}
\put(100,30){\vector(1,0){50}}
\put(200,30){\vector(1,0){140}}
\put(120,20){$l_i=506$}
\put(250,20){$l_j=1384$}
\put(175,50){No Prediction}
\linethickness{0.1mm}
\put(127,40){\line(1,0){93}}
\put(127,30){\line(0,1){10}}
\put(90,35){$o_i=275$}
\put(220,30){\line(0,1){10}}
\put(225,35){$o_j=194$}
\end{picture}
\end{center}
\end {figure}
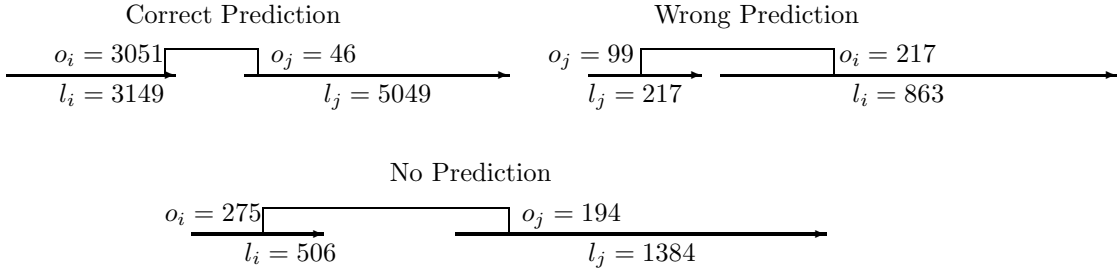

It is also instructive to consider examples from real data. We show 
three cases from a real data set: one in which SLIQ made a
correct prediction, one in which SLIQ made a wrong prediction and one where 
SLIQ did not make any predictions (Fig. \ref{real_exm}). 
We explain precisely which inequalities are violated in the figure caption.
The real examples show the difficulties of making SLIQ predictions when the 
reads fall close to the center of a contig or when the contig lengths are
small relative to the insert size.

\subsection{Naive Scaffolding Algorithm}
The contig digraph constructed in Algorithm \ref{alg1} can be directly
processed to build linear scaffolds. To illustrate this point, 
here we present a naive scaffolding algorithm (Algorithm \ref{naive_alg}).

\begin{algorithm}
\caption{ Naive Scaffolder}
\label{naive_alg}
\begin{algorithmic}[1]
\STATE $G(V,E)=$Construct Contig Digraph (Algorithm \ref{alg1})
\STATE Identify and remove junctions from $G$. Junctions are defined as
articulation nodes with degree $\geq$ 3 that connect at least 3 subgraphs 
of $G$ of size larger than some given threshold. The size 
of a subgraph is defined as the sum of all contig sizes in that subgraph.
\STATE Identify all simple cycles in $G$ and remove the 
edge with the lowest weight from each simple cycle.
\STATE If $G$ still contains strongly connected components, those components 
are removed. $G$ is now a directed acyclic graph.
\STATE Output each weakly connected component of $G$ as a separate 
scaffold.
\STATE The order of contigs in each scaffold is computed by taking 
the topological ordering of the nodes of their respective weakly connected
component in $G$.
\end{algorithmic}
\end{algorithm}

To analyze the computational complexity of the naive
scaffolding algorithm, let $N$ be the number of MPRs in the  
library. Constructing $G$ takes $O(N)$ time. Finding
articulation points takes $O(n+m)$ time where $n=|V|$ and $m=|E|$ 
\cite{articulater}. If we have $a$ articulation nodes,
then finding junctions takes $O(an)$ time. Identifying and breaking simple 
cycles takes $O((n+m)(c+1))$ time where $c$ is the number of simple cycles 
\cite{sim_cy}. Finally, topological sorting takes $O(n+m)$ time. 
In total the complexity of the naive scaffolding algorithm is 
$O(N)+O(n+m)+O(an)+O((n+m)(c+1))=O(N)+O(an)+O((n+m)(c+1))$. In practical data
sets, $a$ and $c$ are small constants and $N>>n,m$. Thus for practical 
purposes the time complexity of the algorithm is $O(N)$.

\section{Experimental Results}
\begin{table}[ht]
\begin{center}
\caption{Descriptive statistics about the datasets.
$R$ is the read length,
cov is the coverage,
$L$ is the reported insert length,
$L_r$ is the real insert length calculated by mapping reads to the 
reference genome and
$\sigma$ is the standard deviation of $L_r$.}
\label{data_sets}

\begin{tabular}{|c|c|r@{.}l |p{2.25cm}|p{2cm}|c|c|c|c|c|c|}
\hline
Set ID& Organism &  Size& & Ref. Genome & Read Lib & R &cov & L & $L_r$ & $\sigma$  \\
\hline
\hline
PSU & {\it P. suwonensis}& 3&42 Mb & CP002446.1 & SRR097515&76 & 870x & 300 & 188.78 & 18.77 \\
\hline
PSY & {\it P. syringae} & 6&10 Mb &NC\_007005.1 &\cite{psy_data}&36 & 40x & 350 &  384.11 & 67.13 \\
\hline
SY-CE &  {\it C. elegans} & 100&26 Mb &NC\_003279-85 & SRR006878 & 35 & 38x & 200 & 232.13 &  54.44  \\
\hline
PST &  {\it P.stipitis} & 15&40 Mb & \cite{meraculous} & \cite{meraculous} & 75 & 25x &  3.2K & 3.27K &   241.50  \\
\hline
DS &  {\it D.simulans} & 109&69 Mb &NT\_167066.1-68.1, NT\_167061.1, NC\_011088.1-89.1, NC\_005781.1 & SRR121548, SRR121549 & 36 & 62x &  N/A & 187.99 &  61.47  \\
\hline
SY-HS & {\it H.Sapiens} & 3&30 Gb &  NCBI36/ hg18 & ERA015743 & 100 & 45x & 300&  310.63& 20.74 \\
\hline
HS & {\it H.Sapiens} & 3&30 Gb &  NCBI36/ hg19 & ERA015743 & 100 & 45x & 300&  310.63& 20.74 \\
\hline
\end{tabular}
\end{center}
\end{table}

To demonstrate the performance of our algorithms in practice, we ran them
on five real data sets and two synthetic 
data sets. The data sets represent genomes ranging in size from 
small bacterial genomes (3Mb) to large animal genomes (3.3Gb) 
(see Table \ref{data_sets} for details).

For each data set, we obtained a publicly available mate pair library. 
We used publicly available pre-built contigs for the 
{\it Drosophila simulans} (DS) and human (HS) \cite{allpath_human} data sets. Pre-built contigs
were not available for the three microbial data sets --- {\it P. suwonensis}
 (PSU), {\it P. syringae} (PSY) and {\it P. stipitis} (PST) --- 
so we used the short read assembler VELVET~\cite{velvet} to construct 
contigs. All software parameters and sources for the data are provided in 
Table \ref{param_table}. 
For the two synthetic datasets, {\it C. elegans} (SY\_CE) and human (SY\_HS), 
we constructed contigs by mapping reads back to the reference 
genome and declaring high coverage regions to be contigs. We will discuss the 
performance of the algorithms on the synthetic data sets at greater length in
the Discussion. We mapped the reads to the 
contigs using the program Bowtie (v. 0.12.7) \cite{bowtie}. 
Below we only report results for the uniquely mapped reads because we know the 
ground truth for them. 

\subsection{Comparison of SLIQ and Majority Voting Predictions}
On all the real data sets, SLIQ was highly accurate in predicting both 
relative orientation ($>75\%$) and position ($> 80\%$) (Table \ref{edge_results}). 
For orientation prediction,
SLIQ and majority filtering produced almost identical accuracies except for
the case of {\it P. stipitis} (PST) where SLIQ had lower accuracy. 
One possible reason might be that the PST library used long mate pair reads 
which may be more inaccurate than the other libraries we tested. Conversely, 
for PST, majority voting gave far worse accuracy (16.5\%) than SLIQ (75\%) in 
relative position prediction, confirming that this data set is an outlier.
 
\begin{figure}[ht]
\begin{center}
\caption{Comparison of the accuracy of SLIQ and majority voting for relative 
position prediction using that same data shown in Table \ref{edge_results}}
\label{bar_chart}
\includegraphics[scale=0.75]{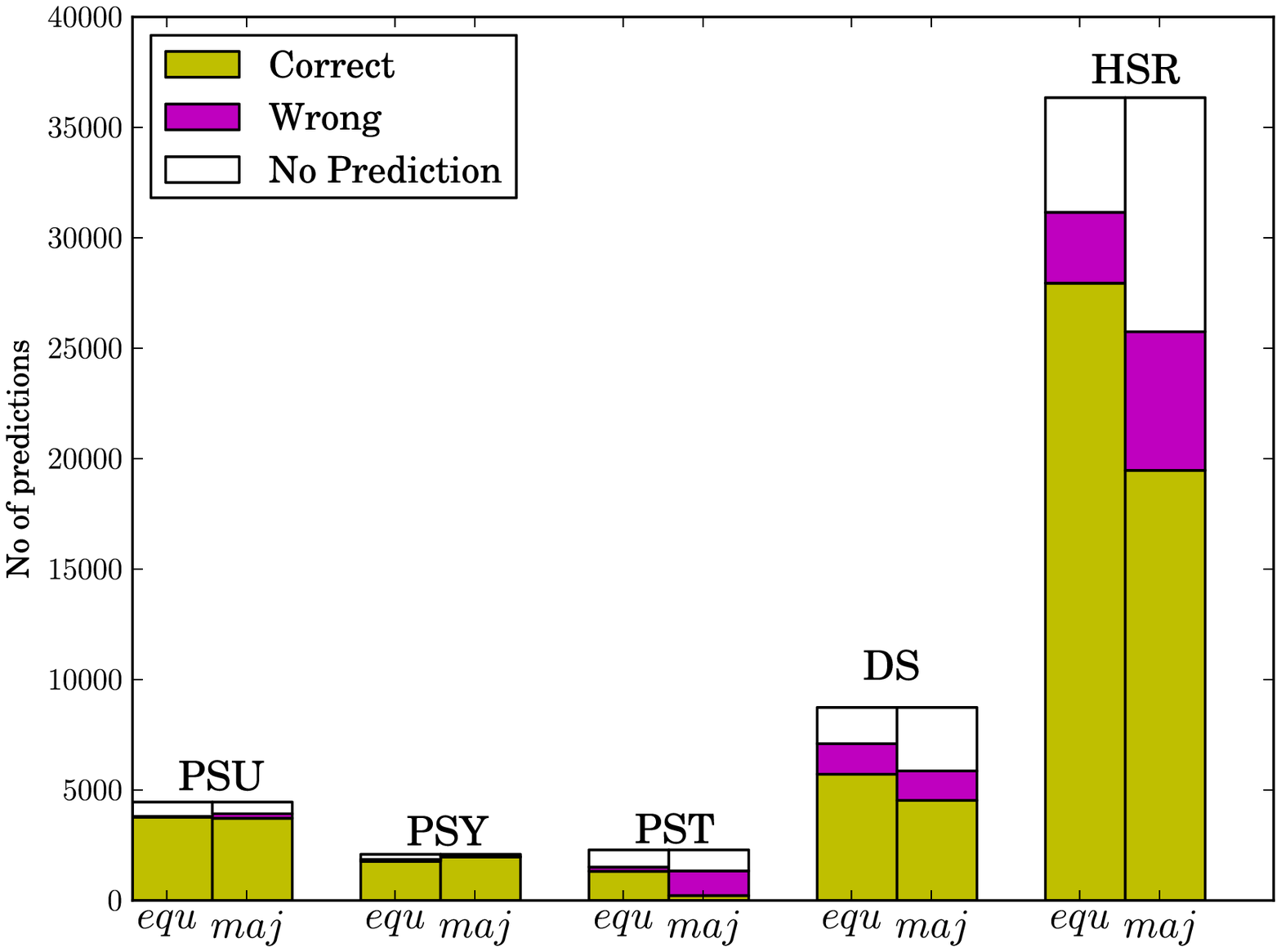}
\end{center}
\end{figure}

Focusing only on the position predictions, SLIQ showed a significant
advantage in both the number and accuracy of the predictions compared to 
majority voting for the more complex genomes --- {\it D. simulans} and human 
(Fig.~\ref{bar_chart}). Importantly, the improvement was particularly large
for the human genome.

Finally, Table \ref{edge_comp_results} gives a more detailed comparison of 
cases where the SLIQ and majority voting predictions disagreed. When
the two methods disagreed, SLIQ clearly outperformed majority voting 
procedure. For example, for human, when the methods disagreed, SLIQ was
right in 1852 cases and majority voting in only 165 cases.
SLIQ was also generally more accurate when considering only the predictions
made uniquely by each method, except in one case (PSY). \\
\begin{table}
\begin{center}
\caption{Summary of the results of SLIQ vs. majority filtering for 
contig graph edges of five real datasets. Here, 
$n$ is the total number of edges connecting two different contigs,
$w_e$ is the minimum wieght of an edge for SLIQ prediction,
$n_o$ is the number of edges for which we can predict relative orientation,
$e_o$ is the accuracy of relative orientation prediction,
$n_p$ is the number of edges for which we can predict relative position,
$e_p$ is the accuracy of relative position prediction and 
$w_m$ is the minimum weight of an edge for majority prediction.
The same notations is used for majority filtering except with prime.
}
\label{edge_results}
\begin{tabular}{|c|c|c|c|c|c|c|c|c|c|c|c|c|c|c|c|c|c|}
\hline
Set ID & $n$ &$w_e$ & $n_o$ & $e_o$ & $n_p$ & $e_p$ & $w_m$ & $n_o'$ & $e_o'$ & $n_p'$ & $e_p'$\\
\hline
\hline
PSU & 4454 &2 & 2507 & 99.69\% & 3803 & 99.21\% & 4 & 3942 & 99.59\% & 3925 & 94.87\% \\
\hline
PSY & 2086 &2 & 1628 & 98.40\% & 1852 & 95.62\% & 4 & 2019 & 98.56\% & 1990 & 98.59\%\\
\hline
PST  & 2291 &1 & 1233 & 75.18\% & 1516 & 87.33\% & 2 & 1365 & 97.87\% & 1336 &  16.54\% \\
\hline
DS  & 8738 &1 & 6305 & 92.18\% & 7097 & 80.55\% & 2 & 6390 & 91.87\% & 5861 &  77.25\% \\
\hline
HS  & 36346 &1 & 31799 & 79.56\% & 31153 & 89.71\% & 2 & 32676 & 79.14\% & 25750 &  75.62\% \\
\hline
\end{tabular}
\end{center}
\end{table}


\begin{table}
\begin{center}
\caption{Comparison of position predictions between
the SLIQ and majority voting methods. Here, 
$n_a$ is the number of predictions where the methods agreed,
$n_d$ is the number of predictions where the methods disagreed,
$n_{d_e}$ is the number of predictions not in agreement where SLIQ was correct,
$n_{d_m}$ is the number of predictions not in agreement where majority voting was correct,
$n_e'$ is the number of predictions made only by SLIQ,
$e_q$ is the accuracy of predictions made only by SLIQ,
$n_m'$ is the number of predictions made only by majority voting,
$e_m$ is the accuracy of predictions made only by majority voting.
}
\label{edge_comp_results}
\begin{tabular}{|c|c|c|c|c|c|c|c|c|}
\hline
Set ID & $n_a$ & $n_d$ & $n_{d_e}$&  $n_{d_m}$  & $n_e'$ & $e_q$ & $n_m'$ & $e_m$\\
\hline
\hline
PSU & 3089 & 646 & 643 &  3  & 68 & 95.58\% & 190 & 90.52\%\\
\hline
PSY & 1519 & 287 & 235 & 52  & 46 & 86.95\% & 184 & 96.19\%\\
\hline
PST  & 290 & 794 & 784 &  10  & 432 & 58.56\% & 252 & 25.00\%\\
\hline
DS  & 2447 & 820 & 804 &  16  & 409& 93.15\% & 2035 & 76.41\%\\
\hline
HS  & 16425 & 2017 & 1852 &  165  & 12711 & 85.67\%& 7308 & 52.73\%\\
\hline
\end{tabular}
\end{center}
\end{table}

\begin{table}
\begin{center}
\caption{Parameter values used in the analysis of all datasets. 
$v$ is the number of mismatches allowed in read mapping (Bowtie v.0.12.7).
}
\label{param_table}
\begin{tabular}{|c|c|p{6cm}|p{6cm}|}
\hline
Data Set & $v$ & contig construction & contig mapping \\
\hline
\hline
PSU & 2 & (velvet) Hash length=21, cov\_cutoff=5, min\_contig\_lgth=150 & (vmatch) min match length $l=150$, Hamming distance $h=0$\\
\hline
PSY & 0 & (velvet) Hash length=21, cov\_cutoff=5, min\_contig\_lgth=150 & (vmatch) min match length $l=150$, Hamming distance $h=0$\\
\hline
PST & 0 & (velvet) Hash length=35, cov\_cutoff= auto, min\_contig\_lgth=100 & (vmatch) min match length $l=200$, Hamming distance $h=5$\\
\hline
SY-CE  & 1 & (synthetic) cov cutoff=5, min contig len=$L$ & available from synthetic construction\\
\hline
DS  & 2 & accession number AASR01000001-AASR01050477 & (vmatch) min match length $l=200$, Hamming distance $h=5$\\
\hline
SY-HS  & 2 & (synthetic) cov cutoff=3, min contig len=$2R$ & available from synthetic construction\\
\hline
HS  & 3 & accession number AEKP01000001:AEKP01231194 & (vmatch) min match length $l=300$, Hamming distance $h=0$\\
\hline
\end{tabular}
\end{center}
\end{table}

\subsection{Computing the Optimal Insert Length}
In our experiments, we found that using a slightly larger value for $L$ than 
that reported or estimated
increased both $n_p$, the number of MPRs for which we could make a relative
position prediction, and $e_p$, the accuracy of relative 
position prediction. This may seem surprising at first given Equation 
\eqref{off_imp}. However, for $n_p$ it can be seen from Fig. \ref{gaps_fig}
that underestimating $L$ would reduce $g_{ij}$ which would lead to more 
overlaps between contigs. Since we assume that the maximum contig overlap is 
$R$, underestimating $L$ would remove many MPRs from the predictions. 
However, at the moment we do not have an explanation for the observed 
increase in $e_p$, the prediction accuracy.

On the other hand, using a slightly smaller value for $L$ increased $n_o$, 
the number of MPRs for which we could make a relative orientation prediction,
while $e_o$, the prediction accuracy for orientation, remained constant. 
We suspect that a lower $L$ makes Equation \eqref{ori1} and \eqref{ori2} harder
to pass and thus less MPRs are excluded by the mutual exclusion test.

\subsection{Computing the Rank of MPRs}
\begin{figure}
\begin{center}
\caption{Change in the prediction accuracy, $e_p$, as we restrict our
analysis to MPRs of higher rank ($c$)}
\label{confidence}
\includegraphics[scale=0.5]{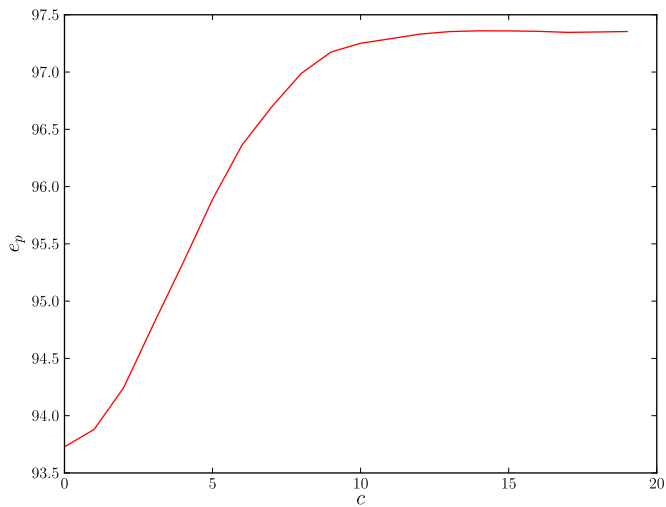}
\end{center}
\end{figure}

Our experimental results also agree with our illustrative cases 
(section \ref{two_cases}) in that the prediction accuracy 
decreases as $2(o_i- o_j)$ gets closer to 
$(l_i- l_j)$ which intuitively means that the reads are falling closer to the 
center of the contigs. To address this issue we can rank the MPRs by the 
minimum value of $c$ for which they fail to pass the more stringent inequality
$|2(o_i-o_j)-(l_i-l_j)|> cR$. We say that an MPR has rank $c$ if and only if $c$ is the 
smallest positive integer such that $|2(o_i-o_j)-(l_i-l_j)|\leq cR$ and
MPRs with higher rank are considered more confident with regards to their 
prediction.
Fig. \ref{confidence} shows how the prediction accuracy depends on the
rank of the MPRs in the PSY dataset. \\

\subsection{Performance of the Naive Scaffolder}
We summarize the results of our naive scaffolder on the five real data sets in
Table \ref{scaf_res} and Table \ref{run_time}. For all data sets, the orientation accuracy was very
high ($> 97\%$) and the position accuracy was also high ($>89\%$). 
While the genome coverages of PSU and DS may appear surprising, note that the 
PSU library had a very high coverage while the DS library had low 
coverage and was also made up of a number of different {\it D. simulans} 
strains. It is likely that the PSU contigs include misassembled fragments
in the contigs, making the total length of the contigs larger than the genome 
size. For DS, the combination of low coverage and relatively high rates of
sequence differences between the different {\it D. simulans} strains likely
resulted in lower genome coverage.\\
\begin{table}[ht]
\begin{center}
\caption{Summary of the results of our naive scaffolder on real data. 
N50 is the length $n$ such that 50\% of bases
are in a scaffold of length at least $n$. The position 
accuracy measures how many neighboring contigs in the scaffold were
placed in the correct order.}
\label{scaf_res}
\begin{tabular}{|c|c|p{2cm}|p{2cm}|p{1.75cm}|}
\hline
Data Set & N50 & Genome Coverage& Orientation Accuracy& Position Accuracy\\
\hline
\hline
PSU & 17K & 116.1\% &  99.64\% & 97.95\%\\
\hline
PSY & 75K  & 90.98\%  & 98.26\% & 93.42\%\\
\hline
PST  & 215K  & 97.89\% &  98.90\% & 89.89\%\\
\hline
DS  & 942  & 59.48\%   & 97.52\% & 96.07\%\\
\hline
HS  & 18k  & 79.27\%   & 98.28\% & 98.03\%\\
\hline
\end{tabular}
\end{center}
\end{table}

\begin{table}
\begin{center}
\caption{Run time comparison of our Naive Scaffolder with two other 
state-of-the-art scaffolders, SOPRA and MIP Scaffolder. 
All times are the sum of the user and system times
reported by the Linux {\tt time} command. We ran all software on a 
48 core Linux server with 256GB of memory. [NOTE to reviewers: MIPS has 
been running for more than 1500 minutes and we will insert
exact running times in the final manuscript]
}
\label{run_time}
\begin{tabular}{|c|c|c|c|}
\hline
Data Set & Naive Scaffolder & SOPRA & MIP Scaffolder \\
\hline
\hline
PSU & 6m40.39s &  237m27.237s & $>$1200m\\ 
\hline
PSY &  59.36s & 44m57.604s & $>$1200m\\
\hline
PST  & 67.21s & 3009m29.224s & $>$1200m\\
\hline
DS  & 7m7.449s & N/A& $>$1200m\\
\hline
HS  & 241m33.928s & N/A & $>$1200m\\
\hline
\end{tabular}
\end{center}
\end{table}
\section{Discussion}
\label{sec:discussion}
In conclusion, we have presented a mathematical approach and an algorithm for constructing a contig
digraph that encodes the relative positions of contigs based on mate pair
read data. Our main insight is the derivation of a set of simple linear 
inequalities derived from the geometry of contigs on the line that we call 
SLIQ. We can use SLIQ both to efficiently filter out unreliable mate-pair 
reads (MPR) and predict the relative positions and orientations between 
contigs. We have shown that SLIQ outperforms the commonly used majority voting 
procedure for the prediction of relative position of contigs while both methods
are very accurate for orientation prediction. The contig
digraph can also be directly processed into a set of linear scaffolds and we 
have presented a simple scaffolding algorithm for doing so. Our naive 
scaffolder has high accuracy on all data sets tested and is very efficient ---
for practical purposes, as it takes time linear in the size of the mate pair 
library and it is also very fast compared to other state-of the art 
scaffolders. The output of our naive scaffolder can either be used directly
as draft scaffolds or used as a reasonable starting point for refinement with more complex
optimization procedures used in other scaffolders.

One interesting and unexpected finding of our experiments was that the
simple majority voting procedure performs very well for predicting the 
relative positions of contigs if the contigs have few errors. This can be 
seen by the performance of the majority voting procedure when using synthetic 
contigs that are not constructed using {\it de novo} assembly tools but 
rather by mapping the reads back to a reference genome and identifying 
regions of high coverage which is expected to produce much higher quality
contigs (Table \ref{syn_results}). This observation suggests a 
novel way to approach the scaffolding problem in which the contig 
builder would output smaller but higher quality contigs and allow the 
scaffolder to handle the remainder of the assembly. We believe this is a 
significant change in philosophy of genome assembly programs to date in 
which during the contig building step, one generally attempts greedily to build contigs 
that are as long as possible. This view point
also differs considerably from previous approaches to scaffolding 
in which the focus was on resolving conflicts between mate pairs that gave
conflicting information about the relative orientation and position of 
contigs. 

\begin{table}
\begin{center}
\caption{Summary of the results of majority prediction for synthetic datasets
for {\it C. elegans} (SY\_CE) and humans (SY\_HS). 
$n$ is the total number of edges connecting two different contigs,
$w_m$ is the minimum weight of an edge for majority prediction,
$n_o$ is the number of edges for which we can predict relative orientation,
$e_o$ is the accuracy in relative orientation prediction,
$n_p$ is the number of edges for which we can predict relative position and
$e_p$ is the accuracy in relative position prediction
}
\label{syn_results}
\begin{tabular}{|c|c|c|c|c|c|c|c|c|c|c|c|c|c|c|c|c|c|}
\hline
Data Set & $n$ & $w_m$ & $n_o$ & $e_o$ & $n_p$ & $e_p$\\
\hline
\hline
SY-CE  &17620 &3 & 17620 & 99.52\% & 17532&  99.85\% \\
\hline
SY-HS  & 878380 & 3 & 878380 & 98.93\% &  868877&  99.47\% \\
\hline
\end{tabular}
\end{center}
\end{table}

Finally, we are exploring several possible extensions of the SLIQ method.
The first extension is to find the optimal value for $L$, the insert length,
so that we optimize the number and accuracy of relative position and orientation predictions.
The second extension is to assign numerical values to the accuracy of 
prediction of MPRs of a particular rank.
Finally, for the multiply mapped MPRs which were not included in the results,
we plan to identify the most likely mapping for the MPR, for example
by using their ranks.
\pagebreak
\bibliography{assembler}

\section{Disclosure Statement}
No competing financial interests exist.
\section{Author Information}
{\bf Rajat S. Roy}\\
Department of Computer Science\\
Rutgers The State University of New Jersey\\
110 Frelinghuysen Rd\\
Piscataway, NJ 08854-8019\\
Email: rajatroy@cs.rutgers.edu\\
Tel: (732) 445-2001 ext 9715\\
\\
{\bf Kevin C. Chen}\\
Department of Genetics \\
BioMaPS Institute for Quantitative Biology\\
Rutgers, The State University of New Jersey\\
145 Bevier Road\\
Piscataway, NJ, 08854\\
Email: kcchen@biology.rutgers.edu\\
Tel: (732) 445-1027 ext 40055\\
\\
{\bf Anirvan M. Sengupta}\\
Department of Physics and Astronomy\\
BioMaPS Institute for Quantitative Biology\\
Rutgers, The State University of New Jersey\\
136 Frelinghuysen Road\\
Piscataway, NJ 08854-8019 USA\\
Email: anirvan@physics.rutgers.edu\\
Tel: (732) 445-3880\\
\\
{\bf Alexander Schliep}\\
Department of Computer Science\\
BioMaPS Institute for Quantitative Biology\\
Rutgers The State University of New Jersey\\
110 Frelinghuysen Rd\\
Piscataway, NJ 08854-8019\\
Email: schliep@cs.rutgers.edu\\
Tel: (732) 445-2001 ext 1166\\
\\

\end{document}